\documentclass[12pt,preprint]{aastex}
\usepackage{color}
\shorttitle{Reddening Law in IRAS~14026$+$4341}
\shortauthors{Jiang et al.}
\begin{document}
\title{Anomalously Steep Reddening Law in Quasars: An Exceptional Example Observed in IRAS~14026$+$4341}
\author{Peng Jiang\altaffilmark{1,2,5}, Hongyan Zhou\altaffilmark{2,1}, Tuo Ji\altaffilmark{2,1}, Xinwen Shu\altaffilmark{1,2}, Wenjuan Liu\altaffilmark{1,2}, Jianguo Wang\altaffilmark{3,4}, Xiaobo Dong\altaffilmark{1,2}, Jinming Bai\altaffilmark{3,4,2}, Huiyuan Wang\altaffilmark{1,2}, and Tinggui Wang\altaffilmark{1,2}}
\altaffiltext{1}{Key Laboratory for Research in Galaxies and Cosmology, The University of Science and Technology of China, Chinese Academy of Sciences, Hefei, Anhui, 230026, China}
\altaffiltext{2}{Polar Research Institute of China, Jinqiao Rd. 451, Shanghai, 200136, China}
\altaffiltext{3}{National Astronomical Observatories/Yunnan Observatory, Chinese Academy of Sciences, P.O. Box 110, Kunming, Yunnan 650011, China}
\altaffiltext{4}{Key Laboratory for the Structure and Evolution of Celestial Objects, Chinese Academy of Sciences, Kunming 650011, China}
\altaffiltext{5}{LAMOST Fellow}
\email{jpaty@mail.ustc.edu.cn}

\begin{abstract}
A fraction of the heavily reddened quasars require a reddening curve which is even steeper than that of the Small Magellanic Cloud.
In this paper, we thoroughly characterize the anomalously steep reddening law in quasars, via an exceptional example observed in
IRAS~14026$+$4341. By comparing the observed spectrum to the quasar composite spectrum, we derive a reddening curve in the rest-frame
wavelength range of 1200~{\AA}--10000~{\AA}. It is featured with a steep rise at wavelengths shorter than 3000~{\AA}, but no
significant reddening at longer wavelengths. The absence of dust reddening in optical continuum is confirmed by the normal broad-line
Balmer decrement (the H$\alpha$/H$\beta$ ratio) in IRAS~14026$+$4341. The anomalous reddening curve can be satisfactorily reproduced
by a dust model containing silicate grains in a power-law size distribution, $dn(a)/da \propto a^{-1.4}$, truncated at a maximum size
$a_{max}=70~{\rm nm}$. The unusual size distribution may be caused by the destruction of large ``stardust" grains by quasar
activities or a different dust formation mechanism (i.e., the in situ formation of dust grains in quasar outflows). It is also possible
that the analogies of the dust grains observed toward the Galactic center is responsible for the steep reddening curve. In addition,
we find that IRAS~14026$+$4341 is a weak emission-line quasar (i.e., PHL 1811 analogies) with heavy dust reddening and blueshifted
broad absorption lines.
\end{abstract}


\section{Introduction}
Quasars are characterized by a blue continuum, thus yielding ultraviolet
excess as a widely-used technique to identify quasars (e.g., Schmidt \& Green 1983).
However, it was found that the radio-selected quasars have
a wider range of optical colors toward red than the quasars selected
purely on basis of optical criteria (e.g., Webster et al. 1995; Brotherton et al. 2001;
Gregg et al. 2002; White et al. 2003). The apparent red quasars are considered to
be the intrinsic blue ones suffering varying amounts of reddening by dust along the line
of sight (e.g., Gaskell et al. 2004). The statistical study on the color of quasars
in the Sloan Digital Sky Survey\footnote{http://www.sdss.org/} (SDSS; York et al. 2000)
firmly demonstrated the existence of a significant population of dust reddened quasars
(Richards et al. 2003). Hopkins et al. (2004) further characterized
the dust reddening of quasars as Small-Magellanic-Cloud-like reddening
(e.g., Pei 1992; Gordon et al. 2003) and pointed out that 2\% of quasars ($z<2.2$)
selected by the main SDSS targeting algorithm (Richards et al. 2002) have moderate
reddening of $E(\bv)>0.1$. Heavily reddened quasars with $E(\bv) \sim 0.5$--1.0 have been
sparsely detected in SDSS (e.g., Hall et al. 2002; Wang et al. 2005; Meusinger et al. 2012).
The recent dedicated surveys of dust-reddened quasars (Glikman et al. 2012;
Fynbo et al. 2013), using optimized selection criteria, expanded the sample in ranges of
redshift and reddening ($0.1\la z\la3.5$, $0.1\la E(\bv) \la1.5$).
Although dust in the intervening absorption line systems can contribute to
the reddening toward quasars (e.g., York et al. 2006; M{\'e}nard et al. 2008;
Wang et al. 2012), reddening is more correlated with the presence of associated
absorption lines (Hopkins et al. 2004; Fynbo et al. 2013). Thus the reddening toward
quasars is mainly caused by dust grains at the quasar redshift, either it is near
the nucleus or more wide spread in the host galaxies.

Most of the statistical studies on the reddening of quasars suggested an
SMC-like reddening law (e.g, Richards et al. 2003; Hopkins et al. 2004; Glikman et al. 2012).
However, other reddening laws have been reported in individual cases. A ``gray" extinction
curve, which is much flatter than the SMC-like curve, has been proposed by different
authors (Maiolino et al. 2001; Gaskell et al. 2004; Czerny et al. 2004; Gaskell et al. 2007),
indicating a preponderance of large grains near the nucleus. The Milky-Way-like extinction
curve, featured with a broad bump around 2175~{\AA} (e.g., Savage \& Mathis 1979),
has been detected in the quasar spectra (e.g., Junkkarinen et al. 2004; Wang et al. 2004;
Srianand et al. 2008; Noterdaeme et al. 2009; Zhou et al. 2010; Jiang et al. 2010a, b;
Jiang et al. 2011). Although the reddening in most of the cases are due to MW-like dust
in the intervening galaxies, four quasars in Jiang et al. (2011) have a prominent 2175~{\AA}
extinction feature at/close to their emission redshifts. In several high redshift quasars
with $z > 4$, the unusual extinction curve (i.e., relatively flat at $>$ 1700~{\AA} and steeply
rising at shorter wavelengths) produced by the supernova dust has been observed
(Maiolino et al. 2004; Gallerani et al. 2010). The overall UV slope of SN-like extinction curve
is slightly flatter than that of SMC-like curve (Todini \& Ferrara 2001;
Bianchi \& Schneider 2007; Hirashita et al. 2008)

Moreover, Hall et al. (2002) reported two extremely reddened broad absorption line (BAL)
quasars in the SDSS early data release with an anomalously steep reddening curve.
That is even steeper than the SMC-like reddening curve at wavelengths shortward of 2000~{\AA}.
A similar reddening curve is also required to interpret the fact that some of the ``UV-deficient"
narrow line Seyfert 1 (NLS1) galaxies showing a UV dip in their SDSS spectra but having Balmer
decrement (measured in the red part of spectra) close to the theoretical value free from reddening
(Zhou et al. 2006). The spectra of these anomalous red quasars are all characterized by a continuum
break around 3000~{\AA}, i.e., the continuum is significantly reddened at wavelengths shortward
of 3000~{\AA}, but has a similar slope with the quasar composite spectrum (Vanden Berk et al. 2001)
at longer wavelengths. Meusinger et al. (2012) identified hundreds of SDSS quasar with unusual
red continua, a fraction of which might fit the anomalously steep reddening scenario. However, it
is hardly possible to further characterize the anomalously steep reddening law using SDSS spectra
due to the relatively narrow wavelength coverage. Recently, an impressive study on
the steep extinction curves of heavily reddened quasars was presented by Fynbo et al. (2013).
They combined the optical spectra and UKIDSS\footnote{The UKIRT Infrared Deep Sky Survey;
http://www.ukidss,org} photometry in NIR to derive the rest-frame UV/optical reddening curves of
the reddened quasars discovered in their red quasar survey. Interestingly, they found that the
SMC-like curve is too shallow to fit the observed reddening curves in the extended wavelength range,
although it provides a good match to the curves in the observed optical range (in the
rest-frame ultraviolet). Therefore, the steep reddening curve is an important complement to the
commonly used reddening laws in quasars, the nature of  which is much less understood.

In this paper, we present an exceptional example of the anomalously steep reddening
curve observed in a more extended wavelength range (from NIR to FUV in the rest frame of quasar)
toward the quasar IRAS~14026$+$4341. The object was identified as a quasar candidate based
on its warm 60 to 25 $\mu$m infrared color observed by IRAS (Kleinmann et al. 1986; de Grijp et al.
1987). Its quasar nature was confirmed with optical spectroscopy and its redshift ($z=0.320$)
was measured by Low et al. (1988). In this work, we adopted the more recent redshift,
$z=0.3227$, determined by using the SDSS spectrum (Schneider et al. 2010).
IRAS~14026$+$4341 is a low-ionization BAL quasar (e.g., Weymann et al. 1991; Voit et al. 1993;
Zhang et al. 2010) showing Mg~{\scriptsize II}, Al~{\scriptsize III} and C~{\scriptsize IV}
broad absorption troughs (Turnshek et al. 1997; Hines et al. 2001).
The quasar is one of the brightest objects in our survey of the SDSS He~{\scriptsize I}$^{*}$
quasar absorption lines (Ji et al. 2013 in preparation). Its SDSS spectrum is remarkable for the
significant He~{\scriptsize I}$^*\lambda$3889 absorptions spreading in a broad velocity range
of $\sim$0--5000 km s$^{-1}$, thus designated as a metastable He~{\scriptsize I} BAL quasar
(Leighly et al. 2011). Here, we focus on its broad band spectral energy distribution
(SED), which is characterized by a continuum break at the rest wavelength of$\sim$3000~{\AA} due
to dust reddening.

This paper is organized as follows. In \S2, we describe data collection and new observations;
in \S3, we construct the SED and derive the reddening law; in \S4, we discuss the Balmer decrement
in optical spectrum, the dust grain model and the blue template for IRAS~14026$+$4341; and a
brief summary is given in \S5.

\section{Archived Data and New Observations}
IRAS~14026$+$4341 was detected on 27 April 1994 during the Two Micron All
Sky Survey (2MASS; Skrutskie et al. 2006) in the $J$, $H$ and $K_s$ bands.
The quasar was also imaged in the SDSS on 11 March 2003 in the $u$, $g$, $r$,
$i$, and $z$ bands (York et al. 2000). We use the SDSS PSF magnitudes in this paper,
since the object is not resolved in SDSS images. Its magnitudes in FUV and NUV bands were
measured by GALEX (Morrissey et al. 2007) on 5 June 2004. The multi-wavelength
photometry of IRAS~14026$+$4341 are summarized in Table 1.

The far-ultraviolet spectrum of IRAS~14026$+$4341, covering a wavelength range
1570~{\AA}--2330~{\AA}, was obtained with the Faint Object Spectrograph (FOS)
aboard the Hubble Space Telescope (HST) by Turnshek et al. (1997) using the
G190H grating on 7 July 1994. Hines et al. (2001) conducted a spectropolarimetric
observation of IRAS~14026$+$4341 with HST/FOS, using four positions of Waveplate B and
the G270H grating, on 11 September 1995. The total flux density spectrum has a wavelength
coverage from 2210~{\AA} to 3300~{\AA}. The fully processed and well calibrated HST/FOS
spectra were retrieved from the HST/FOS Spectral
Atlas\footnote{http://hea-www.harvard.edu/~pgreen/HRCULES.html} compiled by
Kuraszkiewicz et al. (2004). The SDSS spectrum was observed on 20 April 2004,
providing a wavelength coverage in optical ($\lambda\sim$ 3800~{\AA}--9200~{\AA};
York et al. 2000). We extracted the spectrum from the SDSS data release 7
(DR7; Abazajian et al. 2009).

In order to fill the wavelength gap between the HST/FOS/G270H spectrum and SDSS spectrum,
we observed IRAS~14026$+$4341 using the Yunnan Faint Object Spectrograph and Camera (YFOSC)
mounted on the Lijiang GMG 2.4m telescope on 15 April 2011. The G14 (600 $mm^{-1}$) grating
provides a wavelength range of 3200~{\AA}--7800~{\AA} and a resolution $R\sim 1300$.
The 1.8\arcsec~slit was adopted in the night and two exposures with length of 1800s were obtained.
The raw data was reduced by following the standard IRAF\footnote{http://iraf.noao.edu/} routines.

On 2012 April 16th, we observed the IRAS~14026$+$4341 using the TripleSpec (Wilson et al. 2004)
on the 200-inch Hale telescope to explore the expected strong broad He~{\scriptsize I}$^*\lambda$10830
absorption line. Four 300s exposures were taken in an A-B-B-A dithering mode, with the primary
configuration of the instrument (a spectral resolution of $R \sim 3500$ in a 1.1\arcsec~slit).
Two telluric standard stars were taken quasi-simultaneously. The data was reduced with the
{\it Triplespectool} package, which is a modified version of {\it Spextool} (Cushing et al. 2004).
We present the NIR spectrum here to construct a broad SED for reddening analysis.
The study on He~{\scriptsize I}$^*\lambda$10830 absorption line is beyond the scope
of this paper, but will appear in the paper on our sample of quasar He~{\scriptsize I}$^*$
absorption lines (Ji et al. 2013 in preparation). The details of spectroscopic observations
are presented in Table 2.

In summary, we collected photometric and spectroscopic data of IRAS~14026$+$4341
in the wavelength range between 1500~{\AA} and 24000~{\AA}, corresponding to
$\lambda\sim$ 1130~{\AA}--18140~{\AA} in the rest frame of quasar.

\section{Results}
\subsection{SED: A Continuum Break}
In order to study the SED emitted by nucleus, we first need to estimate
the contamination from the quasar host galaxy. High spatial resolution images of
IRAS~14026$+$4341 have been obtained with the Wide Field Planetary Camera 2 (WFPC2)
aboard HST by Hutchings \& Morris (1995). The host galaxy is categorize as an
elliptical undergoing strong interaction and has an angular size of $\sim$ 4\arcsec.
Hamilton et al. (2002) measured the brightness of host galaxy by subtracting
the nucleus in the image observed using F702W filter (giving a passband of
$\lambda\sim$ 6200~{\AA}--7680~{\AA}), yielding a brightness ratio of host galaxy to
nucleus $\sim$12\%. Since galaxies generally have redder color than that of quasar
in the optical band, the galactic contamination to a quasar spectrum could lead an apparent 
smaller spectral index $\alpha$ (defined as $f_\lambda \propto \lambda^{-\alpha}$) than
the intrinsic one of the quasar.
We estimate the variation of quasar spectral index caused by contamination from host galaxy using
simulation. We mix a quasar composite spectrum having $\alpha \sim 1.5$ in the wavelength
range between 3000~{\AA} and 10000~{\AA} (Glikman et al. 2006)\footnote{The quasar composite
spectrum in Glikman et al. (2006) is less contaminated by host galaxies than that in
Vanden Berk et al. (2001). We refer the readers to Fynbo et al. (2013) for a direct
comparison.} and a galaxy template spectrum in the same flux ratio as IRAS~14026$+$4341
to produce a simulated spectrum. We then measure the new spectral index by fitting a power-law
to the simulated spectrum. All galaxy templates in Rowan-Robinson et al. (2008) are considered
in the simulation. Finally, we find a maximum variation of spectral index $\Delta\alpha = 0.2$ among
the simulated spectra. It is roughly equal to the standard deviation of quasar spectral indecs in
SDSS (Richards et al. 2003; Wang et al. 2012). Therefore, we conclude that the contamination
from host galaxy in the quasar SED is negligible in the rest wavelength range of
3000~{\AA}--10000~{\AA} for IRAS~14026$+$4341. At longer wavelengths, the reddening by dust is
not important, thus we do not take efforts to study the variation of spectral slope in that range.

In the UV bands ($\lambda < 3000$~{\AA}), the spectrum of the quasar IRAS~14026$+$4341 is heavily
reddened. It is nearly impossible to study the contamination from host galaxy in the same manner
for the optical bands. This is because we do not know whether the host galaxy suffers the same
reddening as the nucleus. If the dust grains are in the circumnuclear regions of host galaxy,
the radiation from host galaxy might not be reddened as much as that from the nucleus. Instead,
we estimate the contamination by measuring the residual flux level of the broad C~{\scriptsize IV}
absorption trough. It yields a flux ratio of host galaxy to nucleus $<$10\%. Although the residual
flux might have other explanations (e.g., partial coverage of absorbing gas, scattering photons),
the estimation method can give an upper limit of the flux from host galaxy robustly. Therefore, the SED
of IRAS~14026$+$434 in UV is dominated by the nucleus as well.

We correct all photometric and spectroscopic data for the Galactic extinction using the updated
dust map of Schlafly \& Finkbeiner (2011). As shown in Table 2, the spectra at different wavelengths
were taken using varying slit size in different observation conditions (space-based and ground-based).
The absolute spectral fluxes might be not precise enough is some cases. Thus we have to scale
the observed spectral fluxes to fit the photometric measurements for a recalibration. The HST/FOS
spectra and the SDSS spectrum are well consistent with the photometric measurements in their wavelength
range, while the GMG/YFOSC and P200/TripleSpec absolute fluxes need a scale factor close to unity (within 15\%).

Finally, the rest-frame SED from NIR to FUV is constructed by combine all the observations for
IRAS~14026$+$434 (Figure 1). The SED is dominated by the active nucleus as stated above.
A rescaled quasar composite spectrum (Zhou et al. 2010), which is obtained by combining the SDSS composite
($\lambda <$ 3000~{\AA}; Vanden Berk et al. 2001) and the near infrared template
($\lambda >$ 3000~{\AA}; Glikman et al. 2006), is overplotted in Figure 1 for a comparison.
The composite spectrum was normalized to the SED of IRAS~14026$+$434 around 1$\mu$m.
The SED of IRAS~14026$+$434 is characterized by a continuum break around 3000~{\AA}.
The red-ward part of the observed SED ($\lambda \ga$ 3000~{\AA}) is consistent with the composite spectrum,
showing no significant reddening. However, flux in the blue-ward part ($\lambda \la$ 3000~{\AA}) decreases
steeply toward shorter wavelengths.

\subsection{Anomalously Steep Reddening Law}
In the MW, extinction curves are derived by comparing the fluxes of a reddened star and an
unreddened star having a same spectral type (e.g., Cardelli et al. 1989; Fitzpatrick \& Massa 1990).
For quasars, the composite spectrum is usually assumed to be the intrinsic unreddened spectrum
and the extinction curve can be derived by comparing the reddened quasar spectrum with the composite
one (e.g., Wang et al. 2004; Glikman et al. 2004; Srianand et al. 2008; Noterdaeme et al. 2009;
Zhou et al. 2010; Jiang et al. 2010a; Wang et al. 2012; Fynbo et al. 2011, 2013). However,
the derived extinction curve in this manner is usually a reddening curve, since the normalization
of the composite quasar spectrum is arbitrary.

For IRAS~14026$+$434, we compared the rest-frame SED ($S(\lambda)$) with the rescaled composite
spectrum ($C(\lambda)$) mentioned in \S3.1 and calculated the reddening curve as
$E(\lambda - 1\mu{\rm m})=2.5\log{\frac{C(\lambda)}{S(\lambda)}}$. The prominent emission features
on the observed reddening curve (see Figure 2) are due to the BAL troughs in the observed spectrum.
The mismatch of the broad emission lines in the observed spectrum and the composite spectrum is
responsible for the coarseness/spikes in the reddening curve. We initially attempted to
fit the observed reddening curve with simple functions but failed. Instead, we obtained a smooth
reddening curve using a cubic spline curve. During the fitting procedure, the region of BAL troughs
are blocked, and a boxcar filter in size of 150~{\AA} is used to smooth the observed reddening curve.
We fit jointly the spectroscopic and photometric data with the latter weighted by the band widths.
The derived reddening curve at wavelength $<1\mu$m is tabulated in Table 3 and exhibited in
Figure 2 (orange line). Given that there is no significant reddening redward of
$\sim$4000~{\AA} in IRAS~14026$+$434, the derived reddening curve could be a good approximation to
the real extinction curve (i.e., $E(\lambda - 1\mu{\rm m}) \approx E(\lambda - \infty) = A(\lambda)$)
in this case.

We present three SMC-like reddening curves (Pei 1992) with different color excess in Figure 2
for comparison. Apparently, the reddening curve in IRAS~14026$+$434 has a much steeper slope in UV
than that of SMC-like curve. Although the steep rise in UV can be modeled with SMC-like
curves having moderate E(\bv)$\ga0.25$, the flat slope in
rest-frame $V$, $B$ and $U$ bands cannot be fitted simultaneously with these SMC-like curves.
Since the SMC-like extinction curve has the steepest UV rise among the usually observed extinction
curves, i.e., MW-like curves (Fitzpatrick \& Massa 2007); Large-Magellanic-Cloud-like curves
(Gordon et al. 2002); Starburst-Galaxy-like curves (Calzetti et al. 2000); and SN-like curves
(Gallerani et al. 2010), one will not expect that those observed extinction curves can fit the
anomalously steep reddening law in IRAS~14026$+$434. 

\section{Discussion}
\subsection{Normal Balmer Decrement}
If IRAS~14026$+$434 is intrinsically much bluer than the typical quasar, an SMC-like reddening law with a
large E(\bv) can fit the continuum break in a sense (Hall et al. 2002). We find that a combination of a
very steep quasar continuum ($f_\lambda \propto \lambda^{-3.8}$) and E(\bv)$=0.6$ can reasonably fit the observed SED
of IRAS~14026$+$434. To break the degeneracy, we measure the broad-line Balmer decrement of IRAS~14026$+$434
(H$\alpha$/H$\beta$; Osterbrock 1989) to check whether its optical continuum suffers heavy dust reddening.

The SDSS spectrum of IRAS~14026$+$434, covering a rest-frame wavelength range of 2800{\AA}--6900{\AA}, is
constituted with a featureless
continuum, Fe~{\scriptsize II} emission multiplets (so-called ``pseudo-continuum") and other broad
emission lines as well as He~{\scriptsize I}$^*\lambda$3889 absorptions, but most of the narrow lines
are absent or very weak. The weakness of narrow lines is a common feature for BAL quasars
(e.g., Weymann et al. 1991; Zhang et al. 2010). In this paper, we focus on the H$\alpha$/H$\beta$ ratio,
thus restricting the fit in the rest-frame wavelength range from 4000{\AA}--6900{\AA}. The nuclear
continuum is modeled with a broken power law, with free indices for the H$\alpha$ and H$\beta$ regions.
The optical Fe~{\scriptsize II} emission is modeled with the empirical template in analytical
form\footnote{The implementation of the template functions in Interactive Data Language (IDL) is available
at http://staff.ustc.edu.cn/\~{ }xbdong/Data\_Release/FeII/Template/ \,.}, which is based on measurements
of I\,Zw\,1 by V\'eron-Cetty et al. (2004); see Dong et al. (2008) for details, and Dong et~al. (2011)
for tests of the Fe~{\scriptsize II} modeling. Emission lines are modeled with one to four Gaussians.
The best-fit model is presented in the top panel of Figure 3. The model fits the spectrum overall.
However, the V{\'e}ron-Cetty et al. (2004) Fe~{\scriptsize II} template does not fit the optical
Fe~{\scriptsize II} emissions in the region of H$\beta$ well. We refit the spectrum in H$\beta$
region using the optical Fe~{\scriptsize II} template in Boroson \& Green (1992) locally. The template
produces a better fit. In the left-bottom panel of Figure 3, we show the continuum and Fe~{\scriptsize II}
subtracted spectrum in the vicinity of H$\beta$. It is clear that the Boroson \& Green (1992) template
(black line) subtraction yields a cleaner H$\beta$ line profile than the V{\'e}ron-Cetty et al. (2004)
template (gray line) subtraction does. The H$\beta$ line is modeled with four broad Gaussians but without
any significant narrow component. The subtracted spectrum in the H$\alpha$ region is presented in the
right-bottom panel of Figure 3. The V{\'e}ron-Cetty et al. (2004) template subtraction results in a clean
and symmetric H$\alpha$ profile. The H$\alpha$ line is modeled with four broad Gaussians and a weak
narrow Gaussian. We fit the two emission lines jointly by forcing a common profile for their broad components.
The best-fit width of broad components is 2553.98$\pm$153 km s$^{-1}$ in FWHM\footnote{Full width at half maximum}.
We derive a broad-line H$\alpha$/H$\beta$ flux ratio of 3.4$\pm$0.3, where the uncertainty is estimated
through the bootstrap approach in Dong et al. (2008).

The broad-line H$\alpha$/H$\beta$ ratio is consistent with the intrinsic ratio derived using a large sample
of blue active galactic nuclei (H$\alpha$/H$\beta =$3.06$\pm$0.21; Dong et al. 2008) and well below the value
required in the blue quasar continuum scenario, H$\alpha$/H$\beta =$5.7 corresponding to E(\bv)$=0.6$. This
result suggests that the optical broad lines and continuum is nearly free of reddening in IRAS~14026$+$4341,
the extremely blue quasar continuum scenario therefore can be ruled out.

\subsection{Dust Grain Model and Its Implication}
The extinction curves observed in the local galaxies can be reproduced using the simple dust model
incorporating both silicate and graphite material consists of two separate grain populations,
one of silicate composition and one of graphite composition (e.g., Mathis et al. 1977;
Weingartner \& Draine 2001a). Since a grain absorbs and scatters light most effectively at the
wavelength comparable to its size $\lambda \sim 2\pi a$, the size distribution of grains determines
the functional dependence of the extinction on the wavelength. A power-law size distribution,
$dn(a)/da \propto a^{-3.5}$ in a size range of $5~{\rm nm} < a < 250~{\rm nm}$ ($n(a)$ is the number
density of grains with size $\le a$), can produce a satisfactory fit to the average MW extinction curve
(Mathis et al. 1977). The sightlines in the SMC that lack the 2175~{\AA} extinction bump can be
reproduced by models that lack small carbonaceous grains (Weingartner \& Draine 2001a).

The anomalously steep reddening law observed in IRAS~14026$+$4341 has no significant 2175~{\AA}
extinction feature, thus suggesting that silicate grain is the predominant population. In addition,
the absence of reddening in optical indicates that the size distribution truncates at a small size.
We roughly estimate the truncating size to be $a \la 50~{\rm nm}$, as significant dust reddening
only appears at wavelengths shorter than 3000~{\AA} in the spectrum of IRAS~14026$+$4341.

Laor \& Draine (1993) calculated the optical properties of ``astronomical silicate" grains (with
a composition like that of olivine) using Mie theory, the Ralyeigh-Gans approximation and geometric
optics. We attempt to fit the observed reddening law with their dust
model\footnote{ftp://ftp.astro.princeton.edu/draine/dust/diel/Sil\_81.gz}. The size distribution
is assumed to be a power-law, $dn(a)/da \propto a^{x}$, truncated at a minimum size $a_{min}$ and
a maximum size $a_{max}$. The fitting procedure is realized with grid search. It yields a
power-law with a relatively flat slope, $dn(a)/da \propto a^{-1.4}$. The maximum grain size
$a_{max}=70~{\rm nm}$ is well constrained, but the minimum size hits the boundary at
$a_{min}=1~{\rm nm}$, which is the smallest grain size in Laor \& Draine (1993)
model. The best-fit model is presented in Figure 4 (green solid line). We illustrate the contributions
from the small size grains (purple dashed line) and large size grains (blue dashed line) as
well. The small grains providing opaque at wavelengths shorter than 2000~{\AA} cause a
extremely steep rise in the reddening curve. The large grains are required to attenuate lights
in the wavelength range of 2000~{\AA}--3000~{\AA}.

It is generally believed that dust grains are originally produced by condensation in the circumstellar
envelope of evolved stars and injected into the interstellar medium (ISM) in galaxies by stellar outflows
(Gehrz 1989). The size and composition of interstellar grains then could be altered by physical processes
in the ISM (Draine 2003). In cold dense molecular clouds, large dust grains can form substantially by
coagulation of finer grains, while in warm neutral/ionized medium, large grains can be efficiently destroyed
by supernova shock waves (e.g., Draine \& Salpeter 1979; McKee 1989; Jones et al. 1994). In the circumnuclear
region of a quasar, the energetic radiation from the active nucleus can keep the ISM warm. Thus, the
large dust grains in this environment can be easily destroyed by shocks and blast waves driven by nearby
supernovas and/or the active nucleus itself. Destruction of dust grains may explain the lack of large grains
in the size distribution of IRAS~14026$+$4341.

Another possible explanation for the anomalous size distribution of grains observed in IRAS~14026$+$4341
is that the dust has a different origin from the ``stardust". Elvis et al. (2004) showed that dust grains
can form and survive in the outflowing gas of quasars and suggested that quasars are an important source
of cosmic dust, especially at high redshifts. As a BAL quasar, the outflowing gas of IRAS~14026$+$4341 is
rightly along the line of sight. It is plausible that we are witnessing the newly forming dust grains in
quasar outflows of IRAS~14026$+$4341. Future high resolution and high signal-to-noise ratio spectroscopy
of the broad absorption lines will enable us to determine the location of the outflowing gas and to evaluate
its astrophysical properties, such as density, temperature, kinematics, ionization parameter, metallicity as
well as dust depletion pattern. Then we can examine whether the outflowing gas of IRAS~14026$+$4341 is
suitable for dust grain condensation and survival.

In MW, it is found that the extinction curve toward the Galactic center is significantly steeper in the
optical and NIR than ones observed in other sightlines (e.g, Sumi 2004, Nishiyama et al. 2008, 2009).
If this is a common feature in all the galaxies, the steep reddening curves observed in quasars may be
explained naturally, as quasars obviously probe the central regions of their host galaxies (Fynbo et al. 2013).
Recently, steep extinction curve has been identified toward a gamma-ray burst afterglow (GRB 080605;
Zafar et al. 2012). Both of the observations imply that steep extinction curves may be not necessarily
associated with quasar activities but caused by the peculiar population of dust grains residing in the
galactic centers.

\subsection{A Reddened Weak Emission-Line BAL Quasar}
We deredden the observed spectrum of IRAS~14026$+$4341 with the derived reddening law, and compared the
emission lines in the dereddened spectrum and that in the quasar composite spectrum (Figure 5).
Apparently, the quasar composite spectrum is not a good blue template for IRAS~14026$+$4341. In the
dereddened UV spectrum, the expected strong C~{\scriptsize IV} and C~{\scriptsize III}] broad emission lines
are hardly detected but the broad Fe~{\scriptsize III} multiplets are much stronger than that in the quasar
composite spectrum. These features suggest that IRAS~14026$+$4341 is a reddened PHL 1811 analogy
(Leighly et al. 2007a; Wu et al. 2011, 2012). We present the observed spectrum
of PHL 1811\footnote{The UV spectra of PHL 1811 were taken using STIS aboard the HST (Proposal 9181;
PI: Karen Leighly). We retrieved the calibrated spectra from the Mikulski Archive for Space Telescopes (MAST).
The spectrum in Figure 5 was produced by combining the spectra observed with G140L and G230L gratings.}
of Leighly et al. (2007a) in Figure 5. It looks quite similar with the dereddened spectrum of IRAS~14026$+$4341,
except the absence of broad absorption troughs. Another observational feature of weak emission-line quasars
(i.e., PHL 1811 analogies) is their X-ray weakness (Leighly 2007b). The XMM-Newton observations
confirmed that IRAS~14026$+$4341 is extremely weak in X-ray (Young et al. 2009). We suggest that IRAS~14026$+$4341
is a weak emission-line quasar with heavy dust reddening and BALs, which is rarely seen in the
current sample (e.g., Collinge et al. 2005; Shemmer et al. 2009; Wu et al. 2012).

\section{Summary}
We construct a broad-band SED of IRAS~14026$+$4341 in the rest-frame wavelength range of
1130~{\AA}--18140~{\AA}. By comparing the observed SED and the quasar composite spectrum,
we derive a reddening curve in the wavelength range between 1200~{\AA} and 10000~{\AA}.
It is characterized by a steep rise at wavelengths shorter than 3000~{\AA}, but no significant
reddening at longer wavelengths. We measure the broad-line Balmer decrement in IRAS~14026$+$4341,
yielding a normal H$\alpha$/H$\beta$ ratio seen in SDSS quasar sample. The extremely blue quasar
continuum scenario therefore can be ruled out. We successfully produce the steep reddening
law with a grain model containing silicate grains in a power-law size distribution,
$dn(a)/da \propto a^{-1.4}$, truncated at a maximum size $a_{max}=70~{\rm nm}$.
The distribution lacking large grains is very different from those usually observed in the nearby
galaxies. We propose three possible origins for the anomalous size distribution: (1) modification
of ``stardust" grains by quasars; (2) newly forming grains in quasar outflows; (3) analogies of
the dust grains residing in the Galactic central regions. In addition, we find that IRAS~14026$+$4341
is a weak emission-line quasar (i.e., PHL 1811 analogies) with heavy dust reddening and blueshifted
BALs, which is rarely seen in the current sample of weak emission-line quasars.

\acknowledgements
The authors appreciate the enlightening suggestions from the
anonymous referee, which helped improvement of the quality of this paper.
This project was partially supported by Natural Science Foundation of China
with grants NSFC 11233002, NSFC 11203022, NSFC 11033007, and the SOC program
CHINARE2012-02-03. P.J. acknowledges support from the Fundamental Research
Funds for the Central Universities and China Postdoctoral Science Foundation.

This research uses data obtained through the Telescope Access Program (TAP), which is
funded by the National Astronomical Observatories, Chinese Academy of Sciences, and
the Special Fund for Astronomy from the Ministry of Finance.
Observations obtained with the Hale Telescope at Palomar Observatory were
obtained as part of an agreement between the National Astronomical Observatories,
Chinese Academy of Sciences, and the California Institute of Technology.

Some of the data presented
in this paper were obtained from the Mikulski Archive for Space Telescopes (MAST). STScI
is operated by the Association of Universities for Research in Astronomy, Inc., under
NASA contract NAS5-26555. Support for MAST for non-HST data is provided by the NASA
Office of Space Science via grant NNX09AF08G and by other grants and contracts.
This publication makes use of data products from the Two Micron All Sky Survey,
which is a joint project of the University of Massachusetts and the Infrared Processing
and Analysis Center/California Institute of Technology, funded by the National Aeronautics
and Space Administration and the National Science Foundation.
This research has made use of the NASA/IPAC Extragalactic Database (NED) which is operated
by the Jet Propulsion Laboratory, California Institute of Technology, under contract with
the National Aeronautics and Space Administration. 

Funding for the SDSS and SDSS-II has been provided by the Alfred P. Sloan Foundation,
the Participating Institutions, the National Science Foundation, the U.S. Department
of Energy, the National Aeronautics and Space Administration, the Japanese Monbukagakusho,
the Max Planck Society, and the Higher Education Funding Council for England. The SDSS
Web Site is http://www.sdss.org/.

The SDSS is managed by the Astrophysical Research Consortium for the Participating
Institutions. The Participating Institutions are the American Museum of Natural History,
Astrophysical Institute Potsdam, University of Basel, University of Cambridge, Case
Western Reserve University, University of Chicago, Drexel University, Fermilab, the
Institute for Advanced Study, the Japan Participation Group, Johns Hopkins University,
the Joint Institute for Nuclear Astrophysics, the Kavli Institute for Particle
Astrophysics and Cosmology, the Korean Scientist Group, the Chinese Academy of Sciences
(LAMOST), Los Alamos National Laboratory, the Max-Planck-Institute for Astronomy (MPIA),
the Max-Planck-Institute for Astrophysics (MPA), New Mexico State University, Ohio State
University, University of Pittsburgh, University of Portsmouth, Princeton University,
the United States Naval Observatory, and the University of Washington.

\begin{figure}
\epsscale{1.0}
\plotone{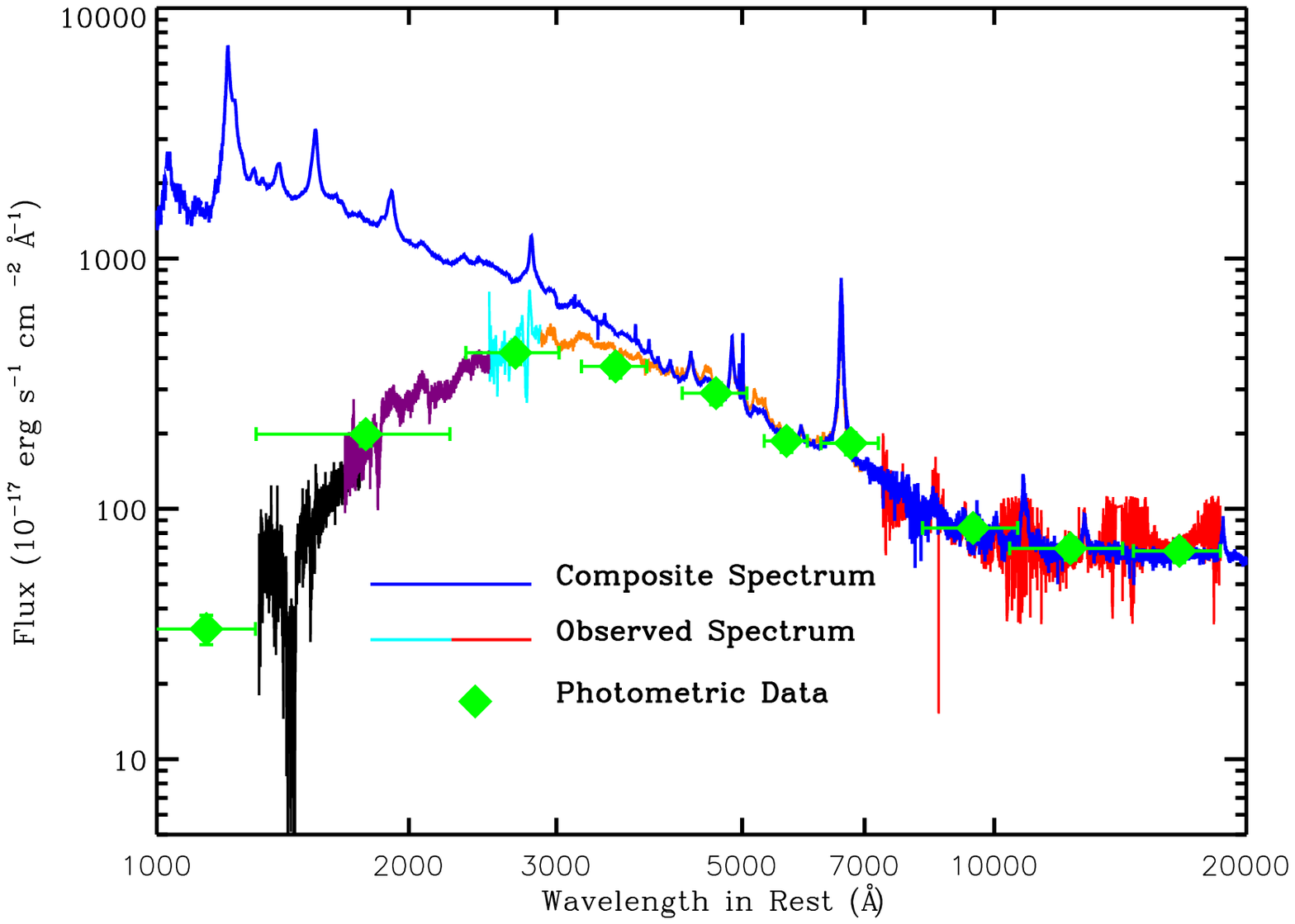}
\caption{\label{fig1} The SED of IRAS~14026$+$4341 in the rest-frame wavelength range of
1130~{\AA}--18140~{\AA}. The green diamonds are the photometric measurements. The observed
spectra are presented in colors: HST spectra in black and purple; YFOSC spectrum in cyan;
SDSS spectrum in orange; and P200 spectrum in red. The quasar composite spectrum is shown
in blue.}
\end{figure}

\begin{figure}
\epsscale{1.0}
\plotone{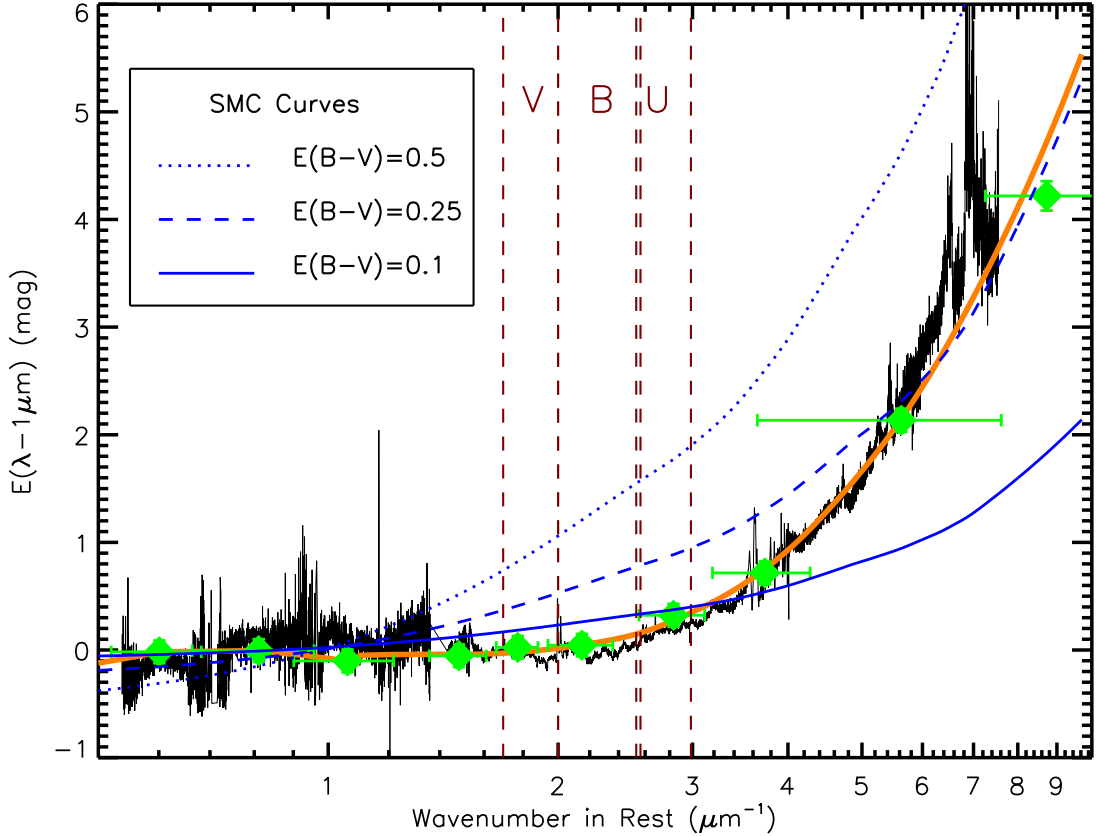}
\caption{\label{fig2} The derivation of reddening law for IRAS~14026$+$4341. The black spectrum
is the observed reddening curve derived with a direct comparison of the SED and the quasar
composite spectrum. The green diamonds are reddening measured in broad-bands. The orange curve
is a smoothed reddening curve by clipping the features and spikes due to the broad absorption
lines and mismatch of emission lines in the spectrum of IRAS~14026$+$4341. Three SMC reddening
curves are presented for comparison (in blue). The wavelength windows of $V$, $B$, $U$ bands are marked.}
\end{figure}

\begin{figure}
\epsscale{1.0}
\plotone{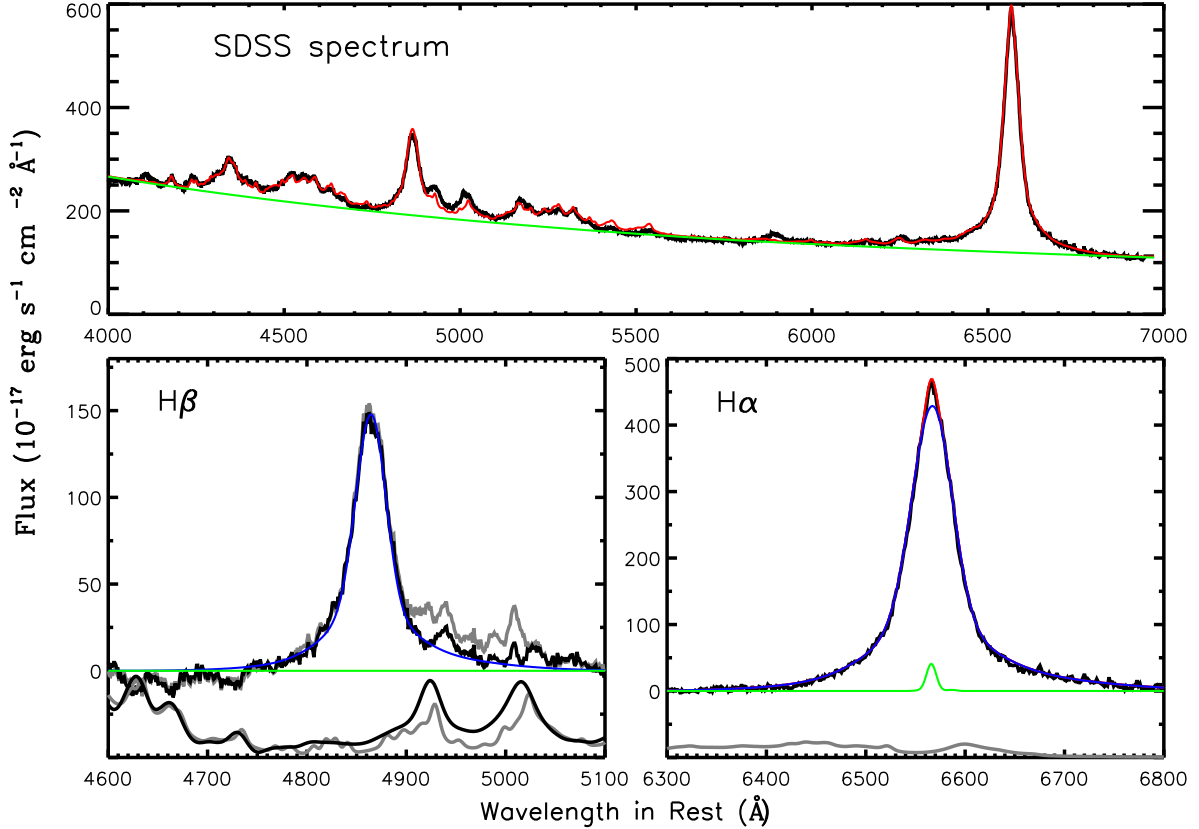}
\caption{\label{fig3} The measurement of Balmer decrement (the H$\alpha$/H$\beta$ ratio)
in SDSS spectrum. The overall spectrum is modeled (in red) with a power-law continuum
(in green), the V{\'e}ron-Cetty et al. (2004) Fe~{\scriptsize II} template and Gaussians
in the top panel. The left-bottom panel shows the continuum and Fe~{\scriptsize II}
subtracted spectrum in the vicinity of H$\beta$. The Boroson \& Green (1992)
Fe~{\scriptsize II} template (in black) produces a better subtraction than the V{\'e}ron-Cetty
et al. (2004) template (in gray). No significant narrow H$\beta$ emission line is detected.
The right-bottom panel shows the subtracted spectrum in the vicinity of H$\alpha$.
The V{\'e}ron-Cetty et al. (2004) template gives a satisfactory fit to Fe~{\scriptsize II}
emission in this region. A narrow component (in green) is required to fit the line profile of
H$\alpha$. The blue curves are the profiles of broad H$\beta$ and H$\alpha$ lines.}
\end{figure}

\begin{figure}
\epsscale{1.0}
\plotone{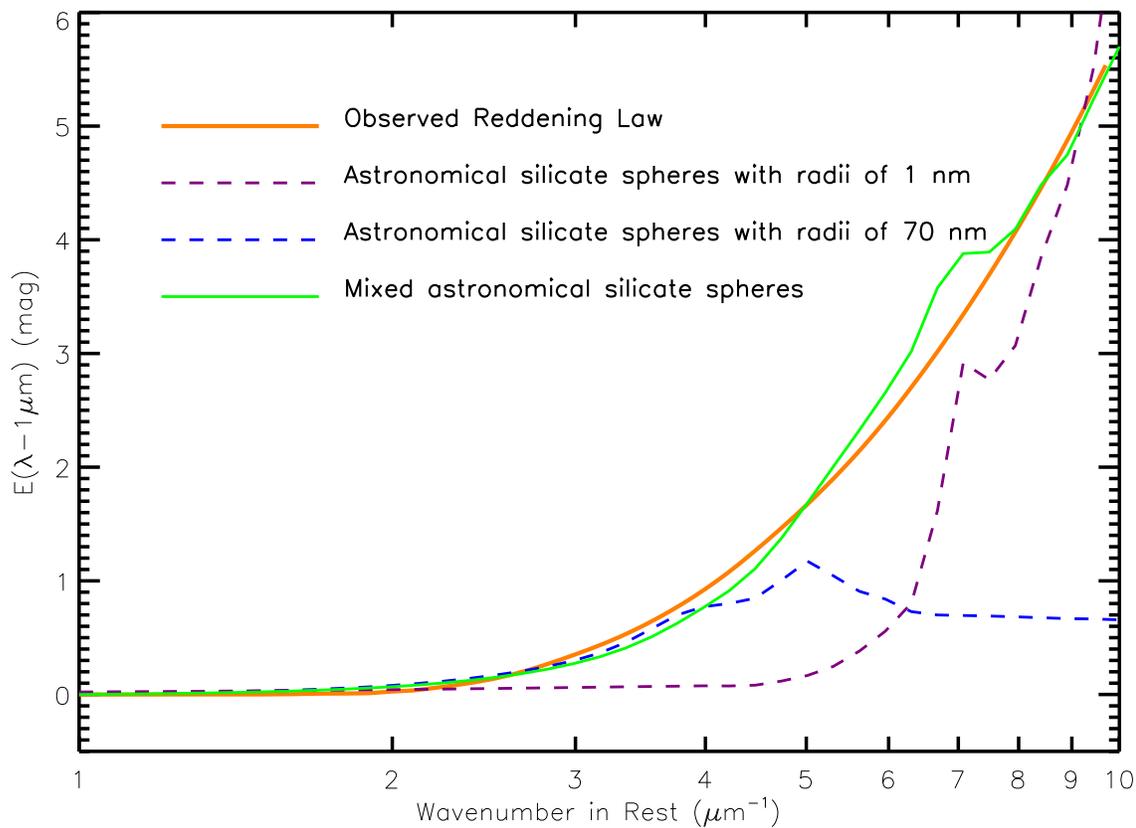}
\caption{\label{fig4} The theoretical reddening curve. The orange curve is the smoothed
reddening curve in IRAS~14026$+$4341. The green curve is produced by a grain model containing
silicate grains (Laor \& Draine 1993) in a power-law size distribution, $dn(a)/da \propto a^{-1.4}$,
truncated at a maximum size $a_{max}=70~{\rm nm}$. The purple dashed curve shows the contribution
from small grains ($a=1~{\rm nm}$) and the blue curve from large grains ($a=70~{\rm nm}$).}
\end{figure}

\begin{figure}
\epsscale{1.0}
\plotone{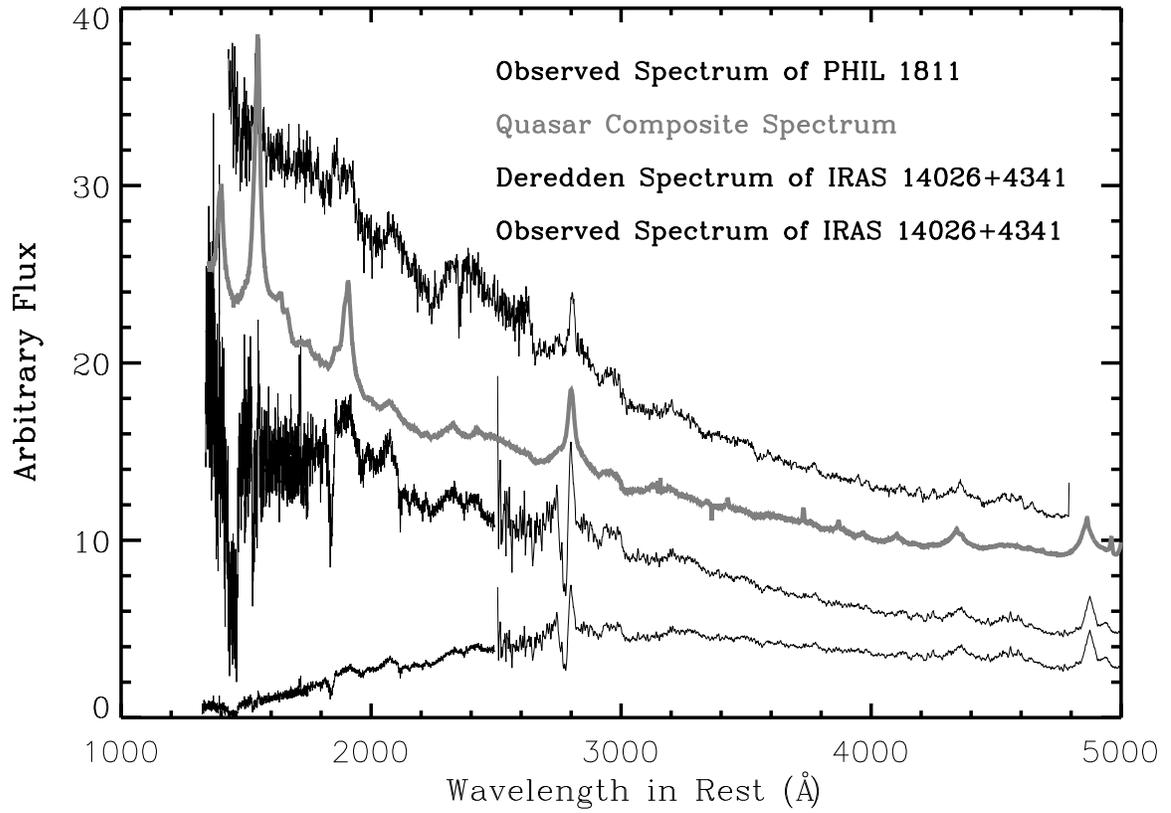}
\caption{\label{fig5} Comparison of deredden spectrum of IRAS~14026$+$4341 with quasar composite
spectrum (gray) and the spectrum of PHIL 1811 (the one on top). IRAS~14026$+$4341
is a weak emission-line quasar with heavy dust reddening and broad absorption lines.}
\end{figure}

\clearpage

\begin{deluxetable}{ccclc}
\tabletypesize{\normalsize}
\tablecaption{Photometric Data\label{tbl-1}}
\tablewidth{0pt}
\tablehead{
\colhead{Band} & \colhead{Value} & \colhead{Facility} & \colhead{Date} &\colhead{Reference}\\
\colhead{ } & \colhead{(mag)} & \colhead{ } & \colhead{(UT)} & \colhead{ }
}
\startdata
FUV & 20.82$\pm$0.32 & GALEX & 2004-06-05 & 1 \\
NUV & 17.99$\pm$0.05 & GALEX & 2004-06-05 & 1 \\
$u$ & 16.19$\pm$0.02 & SDSS &  2003-03-11 & 2, 3 \\
$g$ & 15.67$\pm$0.03 & SDSS &  2003-03-11 & 2, 3 \\
$r$ & 15.32$\pm$0.01 & SDSS &  2003-03-11 & 2, 3 \\
$i$ & 15.36$\pm$0.01 & SDSS &  2003-03-11 & 2, 3 \\
$z$ & 14.95$\pm$0.02 & SDSS &  2003-03-11 & 2, 3 \\
$J$ & 14.20$\pm$0.03 & 2MASS & 1999-04-27 & 4 \\
$H$ & 13.34$\pm$0.03 & 2MASS & 1999-04-27 & 4 \\
$K_s$ & 12.24$\pm$0.02 & 2MASS & 1999-04-27 & 4 \\
\enddata
\tablerefs{(1) Morrissey et al. 2007; (2) York et al. 2000; (3) Abazajian et al. 2009; (4) Skrutskie et al. 2006}
\end{deluxetable}

\begin{deluxetable}{lcccllc}
\tabletypesize{\normalsize}
\tablecaption{Spectroscopic Data\label{tbl-2}}
\tablewidth{0pt}
\tablehead{
\colhead{Range} & \colhead{Slit} & \colhead{$\lambda/\Delta\lambda$} & \colhead{Exp. Time} & \colhead{Instrument} & \colhead{Date} & \colhead{Reference} \\
\colhead{(\AA)} & \colhead{(arcsec)} & \colhead{ } & \colhead{(s)} & \colhead{ } & \colhead{(UT)} & \colhead{ }
}
\startdata
1573--2329 & 1.0 & 1300 & 978 & HST/FOS/G190H & 1994-07-17 & 1, 3 \\
2214--3302 & 1.0 & 1400 & 12590 & HST/FOS/G270H/Pol & 1995-09-11 & 2, 3 \\
3200--7800 & 1.5 &1300 & 3600 & GMG/YFOSC/G14 & 2011-04-15 & 4 \\
3800--9200 & 3.0 & 2000 & 2220 & SDSS & 2004-04-20 & 5, 6 \\
10000--24000 & 1.1 & 3500 & 1200 & P200/TripleSpec & 2012-04-16 & 4 \\
\enddata
\tablerefs{(1) Turnshek et al. 1997; (2) Hines et al. 2001; (3) Kuraszkiewicz et al. 2004; (4) this work; (5) York et al. 2000; (6) Abazajian et al. 2009}
\end{deluxetable}

\begin{deluxetable}{cc}
\tabletypesize{\normalsize}
\tablecaption{Reddening Curve\label{tbl-3}}
\tablewidth{0pt}
\tablehead{
\colhead{$\lambda^{-1}$} & \colhead{E($\lambda$-1$\mu$m)} \\
\colhead{($\mu$m$^{-1}$)} & \colhead{(mag)}
}
\startdata
1.0 & 0.00000 \\
1.1 & 0.00020 \\
1.2 & 0.00046 \\
1.3 & 0.00068 \\
1.4 & 0.00082 \\
1.5 & 0.00103 \\
1.6 & 0.00321 \\
1.7 & 0.00736 \\
1.8 & 0.00904 \\
1.9 & 0.01211 \\
2.0 & 0.02643 \\
\enddata
\tablecomments{This table is available in its entirely in a machine-readable form in the online journal.
A portion is shown here for guidance regarding its form and content.}
\end{deluxetable}

\end{document}